\documentclass[copyright,creativecommons]{eptcs}
\providecommand{\event}{LOCOCO 2010} 
\usepackage{breakurl}             
\usepackage{verbatim}
\usepackage[x11names]{xcolor}
\usepackage{graphicx}
\usepackage{subfigure}
\usepackage{xspace}
\usepackage{amsmath}
\usepackage{amsthm}
\usepackage{varioref}
\usepackage{mdwlist}

\renewcommand{\baselinestretch}{0.985}

\newtheorem{theorem}{Theorem}
\newtheorem{lemma}{Lemma}

\title{\textsc{Configen}: A tool for managing configuration options}
\author{Emmanuel Ohayon \qquad\qquad Matthieu Lemerre \qquad\qquad Vincent David
\institute{Real-Time Embedded System Laboratory\\ CEA LIST, France}
\email{\{emmanuel.ohayon, matthieu.lemerre, vincent.david\}@cea.fr}}
\def\titlerunning{\textsc{Configen}: A tool for management of configuration options}
\def\authorrunning{E. Ohayon, M. Lemerre \& V. David}

\begin{document}

\maketitle{}

\newenvironment{org}{\verbatim ORG ORG}{FIN ORG\endverbatim}
\def\configen{\textsc{Configen}}
\def\code#1{\texttt{#1}}
\def\xxxemmanuel#1{\textcolor{red}{Pour Emmanuel: #1}}
\def\xxxquestion#1{\textcolor{green}{#1}}
\def\xxx#1{\textcolor{red}{#1}}
\def\xxxpaslaplace#1{}
\def\hrulespace{\smallskip\hrule\smallskip}

\def\longerpaper#1{}
\newenvironment{verbatimcode}{\hrulespace
  \begin{minipage}[htbp]{1.0\linewidth}
\verbatim}{\endverbatim\end{minipage}\hrulespace}

\begin{abstract}
  This paper introduces \configen{}, a tool that helps modularizing software.
  \configen{} allows the developer to select a set of elementary components for
  his software through an interactive interface. Configuration files for use by
  C/assembly code and Makefiles are then automatically generated, and we
  successfully used it as a helper tool for complex system software refactoring.
  \configen{} is based on propositional logic, and its implementation faces hard
  theoretical problems.
\end{abstract}
\section{Introduction}

A good way to build secure systems is the top-down approach, where each step
refines the software towards the final implementation. The result is
well-integrated, but quite monolithic. Consequently, further extensions often
lead to an overuse of preprocessor conditionals and some code duplication. It
is then important to refactor and modularize the code, with the goal of
increasing maintainability and code reuse.

We are trying to apply this process to the implementation of the
OASIS~\cite{bib:oasis} kernel, an execution support for hard
real-time safety critical applications. Modularizing this software has
specific requirements. First, the configuration has to be chosen at
compile-time (in particular, qualification for use in safety-critical
environments requires that no dead code remains in the system).
Second, modularity should not impact the degree of performance, in
terms of execution time and memory footprint (for instance,
modularity should not imply new indirections, like C++ virtual
method tables). Thus, the tool should allow the static selection of a subset of
the code in order to implement a specific behavior.

\configen{} is the tool we built to that end. It is composed of two main parts.
The first one is an interactive tool that helps selecting correct software
options with respect to the dependencies between the modules, and is based on
propositional logic. The second part builds the source code following the set of
selected options.

 
The paper is divided as follow. Section~\ref{sec:configen-approach}
explains the concepts and goals of \configen{}.
Section~\ref{sec:usage-patterns-good} provides a set of good practice
rules with concrete examples on how to use \configen{}, as well as
our experience using it with the OASIS kernel \cite{bib:oasis}.
Section~\ref{sec:configen-prototype} presents our current prototype,
and the theoretical problems of its core component, the logic solver.
Section~\ref{sec:state-art} presents related works, and section
\ref{sec:conclusion} concludes.


\section{The \configen{} approach}
\label{sec:configen-approach}

\subsection{Configuration options}

\configen{} operates on \emph{configuration options}, rather than on
\emph{modules}. A module is a part of the code which, when associated
with its dependencies, is ``self-containing'', and often has a defined
meaning that depends on the language (e.g. Java classes, ML modules, C
functions and files). Configuration options represent arbitrary pieces
of code, which encompass the notion of modules, and are thus more
general: it can span from several lines of code inside a function to a
large set of modules.

\label{abstract-code-selectionner}

Formally, a configuration option is a couple $(v,s_v)$ of a boolean variable $v$
and a code selector $s_v$. $v$ is true when the functionality is present, and
false otherwise.
The code selector $s_v$ is a function that, given a value of $v$ and a
code $c$ (a sequence of characters), returns a subsequence of $c$. A
concrete implementation of this function is the use of the C
preprocessor to eliminate conditional code (see
section~\ref{sec:gener-configh-file}). Another one is selection in a
Makefile of a subset of the files to compile or link (section
\ref{sec:gener-makefile}).

\configen{} operates on closed systems, i.e. all the code and
configuration options are assumed to be known when the system is
built. This is a requirement of the ``static configuration'' approach.

Once the values for all configuration options $v$ have been chosen, the
configured code can be obtained by applying all the $s_v$ to the
original code.


\subsection{Relations between configuration options}

\subsubsection{Basic operators}
\label{subsubsec:basic-ops}

Two operators are defined to describe all dependency relations between
configuration options in the system:

\begin{itemize}
\item The \emph{dependency} operator, $a \Rightarrow b$, means that the
  configuration option $a$ can only be present if $b$ is present. It
  is equal to the standard boolean logic implication operator (also
  written $\Rightarrow$).

\item The \emph{interface/implementations} operator, written
  $a:i_1|i_2|\ldots|i_n$, means two things:
  \begin{enumerate}
  \item if the \emph{interface} $a$ is false, then all of the
    \emph{implementations} $i_1 \ldots i_n$ are false;
  \item if $a$ is true, exactly one of $i_1,i_2,\ldots,i_n$ is true.
  \end{enumerate}

  The interface/implementation operator can be expressed by the following
  logical formula:
  \[ \left( \lnot a \land \bigwedge_{1 \le k \le n} \lnot i_k \right) \lor 
  \left( a \land \bigvee_{1 \le k \le n} \Big[i_k \land \textstyle\left(\bigwedge_{l
  \ne k} \lnot i_l \right)\Big]\right)\]
\end{itemize}

\xxxpaslaplace{Note: On aurait pu permettre à l'interface d'être à 0
  lorsque l'une de ses implémentations est à 1. Mais le fait de forcer
  l'interface à 1 lorsque l'implémentation est à 1 permet une
  factorisation de contraintes: les dépendances communes à toutes les
  implémentations d'une interface peuvent être spécifiées comme
  dépendance de l'interface, ce qui simplifie le fichier de dépendance
  et sa représentation graphique. De plus, dans l'interface grpahique
  on ne peut pas cliquer sur les implémentations. Forcer l'interface a
  1 quand les implementations sont a 1 permet d'affecter des valeurs
  aux interfaces ``inutilisées'', comme plateforme dans l'exemple.}

In fact, only this last operator is formally needed, because it is functionally
complete\footnote{\label{foot:operators}Proof: $\lnot x$ is $(x:x|x)$, 0 is
$(\lnot x:x|x)$, $x\land y$ is $(0:(\lnot x|\lnot y))$}. But the use of the
dependency operator makes things simpler for both the user and the logic
simplifier.

In our system the complete relationship between the configuration
options can be written as a conjunction of formulas which either use
the interface/implementation operator on several literals, or the
dependency operator on two literals. For convenience, we also allow
the use of the $\land$ operator on the right side of a $\Rightarrow$ operator.
Other operators may be added in the future, but as of now, we do not
believe that $\lnot$ or $\lor$ are useful operators. We believe that
restricting the number of operators is simpler for the developer and
encourages good software practices, as described in
section~\ref{sec:usage-patterns-good}.

\subsubsection{Textual and graphical representation}

\begin{figure}[htbp]
  \centering

\subfigure[Example of textual representation.]{\vbox{\framebox{\hbox{\begin{minipage}[htbp]{0.7\linewidth}
        \begin{flushleft} \ttfamily
sched -> clock \& ctxlist\\
clock: clock\_llsc | clock\_spinlock\\
clock\_spinlock -> spinlock\\
clock\_llsc -> llsc\\
ctxlist: ctxlist\_llsc | ctxlist\_spinlock\\
ctxlist\_llsc -> llsc\\
ctxlist\_spinlock -> spinlock\\
spinlock: spinlock\_ppc | spinlock\_s12xe | spinlock\_llsc\\
spinlock\_llsc -> llsc\\
llsc: llsc\_arm | llsc\_ppc\\
llsc\_arm -> arm\\
llsc\_ppc -> powerpc\\
spinlock\_s12xe -> s12xe\\
spinlock\_ppc -> powerpc\\
plateform: powerpc | s12xe | arm\\
        \end{flushleft}
  \end{minipage}}}\vspace{1mm}}\label{fig:exemple-textual-representation}}

\vspace{0.8cm} { \subfigure[Graphical representation of
  \ref{fig:exemple-textual-representation}.]{\scalebox{0.9}{
    \includegraphics[width=\textwidth]{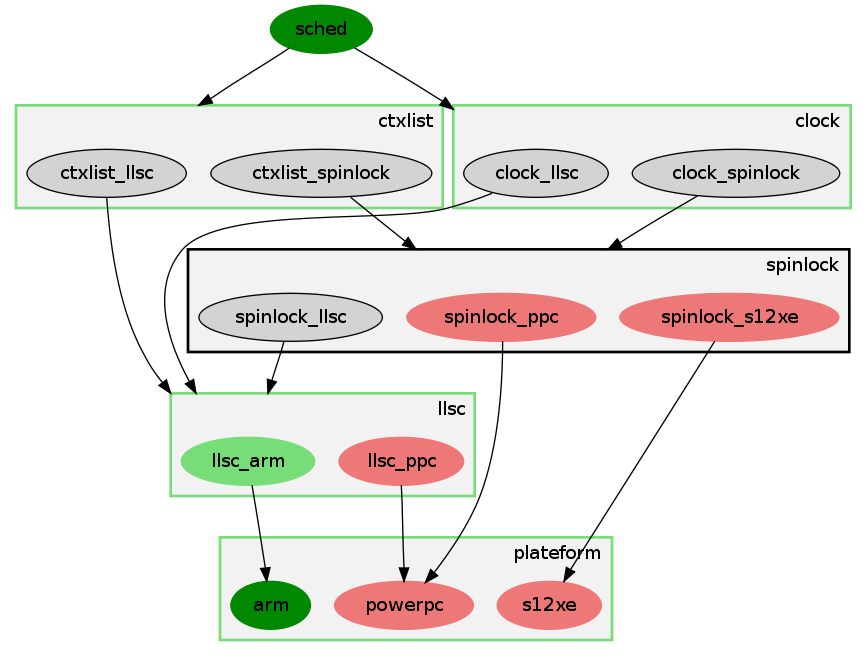}}\label{fig:example-graphical-representation}}
\caption{An example of textual and graphical representation. Each node
  represents a configuration option. Rectangular nodes represent
  interfaces, and the nodes they encompass are their implementations.
  Arrows represent dependencies. Colors represent a partial resolution
  of the logic problem: dark green nodes have been enforced to be
  true, light green ones are deduced true, and light red ones are
  deduced false.}}
  \label{fig:example-spinlock-llsc}
\end{figure}

One of the main interests of using only these two operators is that
they allow nice textual and graphical interfaces. In \configen{}, the
user specifies its dependencies in a special ``\code{deps}'' file,
whose core\footnote{The complete BNF allows some extensions, as seen
  sections \ref{sec:gener-makefile} and
  \ref{sec:opti-behav-small-small-piece-code}.} BNF is simple:

\begin{verbatim}
<deps> ::= { <dep_line> | <iface_line> } *
<dep_line> ::=  <id> "->" <id> { "&" <id> }* "\n"
<iface_line> ::= <id> ":"  <id> { "|" <id> }* "\n"
<id> ::= [a-z][a-z0-9_]*
\end{verbatim}

Each id represents a configuration option (i.e. a boolean variable),
and lines can either express a dependency relation or an
interface/implementation relation. The whole program relation is the
conjunction of the relations in each line.
Figure~\ref{fig:exemple-textual-representation} presents an example of
this textual representation.

The relations between configuration options using our operator also
admits a nice graphical representation. This representation is a graph
where nodes represent configuration options, and arrows dependency
relations. An interface is a box that encloses its implementations. 

Figure~\ref{fig:example-graphical-representation} gives an example on
how the scheduler part of our kernel can be modularized. The
microkernel can run on three different embedded platform, with ARM,
PowerPC, or S12XE processors. ARM and PowerPC both have a LL/SC
(load-linked/store conditional) instruction, S12XE and PowerPC provide
hardware spinlocks. Note that spinlocks can also be implemented using
LL/SC. At last, the scheduler depends on two subsystems, to handle the
current clock and a list of contexts, for which two versions exist: one
that uses spinlocks, and one that uses LL/SC instruction.

Colors represent valuation of boolean variables, as described in the following
section. The interactive solver is
described in details in section \ref{subs:logicsolver}.

\subsection{Tools and integration with the development environment}

\subsubsection{The configuration selector}

A \emph{configuration} is the assignment of a truth value to all
configuration options. The most important requirement for a
configuration is to be correct, i.e. that all the relations between
configuration options are satisfied. It is fairly easy to write a
program that checks that a given configuration is correct.

But such manual writing of a configuration would be tedious for the
user, all the more because our method encourages using many
configuration options (see section~\ref{sec:divid-conf-opti}). Moreover, most
options can be automatically constrained. 

This explains why the configuration selector is necessary. 
Figure~\ref{fig:example-graphical-representation} is a part of a
screenshot of our tool. Its interface is simple: clicking on a
node switches the valuation of the corresponding options between
``enforce true'' (dark green), ``enforce false'' (dark red), and
``unenforced''. Unenforced options can be in different states:
``implied true'' (light green), meaning that all correct
configurations require the option to be true; ``implied false'' (light
red), meaning that all correct configurations require the option to be
false; and ``normal'' (gray), meaning that there exists correct
configuration options where the option is true and others where the
option is false. The tool warns the user when the enforced values are
impossible to satisfy, and allows saving the configuration when every
option has been assigned a value.

In the example, the user has explicitly expressed that he wants the \code{sched}
and \code{arm} configuration options to be true (in dark green). Had the S12XE
platform been selected instead of the ARM one, the configuration would be
\emph{complete}, i.e. every configuration option would have been inferred to be
either true or false.

We found the use of this tool very
intuitive, and that creating a new configuration was fast.

\subsubsection{Generation of a \code{config.h} file}
\label{sec:gener-configh-file}

One of the main use of our tool is the generation of a \code{config.h}
file for use by the C preprocessor. This is the concrete
implementation of the abstract code selector presented
section~\ref{abstract-code-selectionner}.

The \code{config.h} file is generated once all the configuration
options are assigned a value. For every configuration option set to ``true'', a
line \code{\#define CONFIG\_<config option name>} is inserted in this
file.

Every file in the project contains a \code{\#include <config.h>}, and
code can be made optional using \code{\#ifdef CONFIG\_<config option
  name>} or \code{\#ifndef CONFIG\_<config option name>}.

Another advantage of using \configen{} is the assurance that
configuration options are defined consistently. In particular, this
avoids the problem where a conditional is defined only if another
conditional is activated. For instance, the use of spinlock is useful
only on multiprocessor, which can lead to code like this:

\hrulespace
{\hbox{
  \begin{minipage}[htbp]{0.3\linewidth}
\begin{verbatim}
#ifdef CONFIG_SMP
#define CONFIG_SPINLOCK
#endif
\end{verbatim}
  \end{minipage}\hspace{2mm}\vrule\hspace{2mm}
  \begin{minipage}[htbp]{0.4\linewidth}
\begin{verbatim}
#if defined(CONFIG_SMP) && !defined(CONFIG_SPINLOCK)
// conditional code
#endif
\end{verbatim}
  \end{minipage}
}}
\hrulespace\smallskip

\begin{sloppypar}
  Thus the user always has to remind to test the \code{CONFIG\_SMP}
  conditional before testing \code{CONFIG\_SPINLOCK}, which is
  tedious and error-prone. The use of a single, consistent
  \code{config.h} avoids all needs for nested preprocessor
  conditionals.
\end{sloppypar}
With all these problems solved, the use of conditionals in C code becomes
much more readable and maintainable, and allows for reusable code
without sacrificing performance. 

\subsubsection{Generation of \code{Makefile}s}
\label{sec:gener-makefile}

Experience shows that selecting code parts using only preprocessor
conditionals leads to unreadable code. Often, a better way is to
perform a selection of the files to be compiled in the build scripts,
such as Makefiles.

One way to achieve that would be to use conditionals in the Makefile, 
but this makes it harder to read and more error-prone. A better
way is to generate the list of objects and other targets to be built.
To do that, we have extended \configen{} to handle \emph{properties},
which are information attached to the configuration options.
Properties are expressed in the dependency file, as in the following
example:

\medskip
\hrulespace
\begin{minipage}[htbp]{1.0\linewidth}
\begin{verbatim}
ctxlist.objs = ctxlist_common.o
ctxlist_spinlock.objs = ctxlist_spinlock.control.o ctxlist_spinlock.exec.o
microkernel.targets = microkernel
\end{verbatim}
\end{minipage}
\hrulespace
\medskip

These configuration options are used to generate a
file \code{config.mk}:

\medskip
\hrulespace
\begin{minipage}[htbp]{1.0\linewidth}
\begin{verbatim}
all_objs = ctxlist_common.o ctxlist_spinlock.control.o ctxlist_spinlock.exec.o
all_targets = microkernel
\end{verbatim}
\end{minipage}
\smallskip\hrule\smallskip
\medskip

This file is included in the main Makefile for the application:

\medskip
\hrulespace
\begin{minipage}[htbp]{1.0\linewidth}
\begin{verbatim}
all: $(all_targets)
microkernel: $(all_objs)

# Special rules eventually needed to build object files
%.control.o: %.c
      ...
\end{verbatim}
\end{minipage}
\hrulespace

\configen{} can be easily extended to handle new properties, or other build
tools than Makefile.


\xxxpaslaplace{ Autre usage: Usage par un commercial/intégrateur pour
  configurer un logiciel pour son client?}

\section{Usage patterns and good practice}
\label{sec:usage-patterns-good}

Proper use of our tools requires to comply with a set of good
practices, that help writing more modular and understandable code by
using the right amount of configuration options. Indeed if creating
redundant options should be avoided, it is also a bad idea to group
independent concepts into a single configuration option. The following
presents common use cases and the best way to describe them using
\configen{}.

\subsection{Using configuration options for modular construction}

Decomposition into interface and implementations is a common practice
(see the ML module system \xxxpaslaplace{\cite{leroy}}, or C++/Java
abstract and concrete classes). An interface helps understanding the
specification of a module (and how to use it) without needing to
understand its implementation.

In C, defining a function is almost the only way of creating
abstraction, and a function is not enough to define a module. However,
it is possible to write modular software in C by grouping together
several function definitions in one or several files, and grouping all
the functions declarations in one header file that defines the
interface.

\configen{} then helps to make these modules optional, to manage
different implementations of the same module, to state dependencies
between modules, and to automate their build. Moreover, module
dependencies are an important information on how the software is built
and how its modules interact, and \configen{} graphical output is
very useful as a documentation. This helps in making the source code
self-documenting, an important principle for understandable code
(especially in open source software).

\xxxpaslaplace{EXEMPLE UTILE? For instance, one of the OASIS scheduler
  is based on EDF and has to manipulate lists of current contexts
  sorted by increasing deadlines. There are several implementations of
  this scheme: a simple one for uniprocessor scheduling, a parallel
  one with mutex on for embedded bi-core processors (where one core is
  used to perform the scheduling for the other core) [ref patent
  OASIS? pas bon pour cette conf pour l'opensource], and a parallel
  one for embedded bi-core that use llsc.}

\subsection{Optional behavior in small pieces of code}
\label{sec:opti-behav-small-small-piece-code}

There are some configuration options that affect small pieces of code,
typically something too small to write a specific module. For instance,
our scheduler has an optional optimization that requires a small
calculation in order to avoid sending an inter-processor interrupt.
The C code looks like:

\medskip\hrule\smallskip
\begin{minipage}[htbp]{1.0\linewidth}
\begin{verbatim}
#ifdef CONFIG_OPTIMIZE_SEND_IPI
if( do_calculation()) return;
#endif
send_IPI();
\end{verbatim}
\end{minipage}
\smallskip\hrule\medskip

%
%
%
%

The approach we advocate using \configen\ is to define \code{optimize\_send\_ipi}
as an interface with two implementations (\code{optimize\_send\_ipi\_yes} and
\code{optimize\_send\_ipi\_no}), and make \code{sched} depend on it. 
The symbol \code{CONFIG\_OPTIMIZE\_SEND\_IPI} will then be either defined or
``un-defined'', depending on the chosen implementation. For convenience, the
``yes/no'' implementations are automatically declared in the \code{deps} file
when a symbol name ends with a question mark.  The final \code{deps} file is
then:

\medskip\hrule\smallskip
\begin{minipage}[htbp]{1.0\linewidth}
\begin{verbatim}
sched -> optimize_send_ipi?
# (auto) optimize_send_ipi? : optimize_send_ipi_yes | optimize_send_ipi_no
\end{verbatim}
\end{minipage}
\smallskip\hrule\medskip

This kind of optional behavior is not restricted to yes/no choices,
and this scheme accommodates to any number of options.

\xxxpaslaplace{Of course, when an option is always preferable, we could just
  delete the conditional and add a comment in the code saying that
  benchmarks proved that the optimization was worthy.}

\subsection{Optional use of a module}

Sometimes the use of a module can be optional. For instance, when
porting OASIS to a new platform, we do not implement memory protection
in the early stages of development, in order to quickly get to functional
kernel. The recipe in the previous section can be used in this case; basically,
it consists of surrounding all uses of a module by
\code{\#ifdef CONFIG\_USE...}.

This raises a few problems though. First, it leads to many uses of
preprocessor conditionals in the code, which makes it less readable.
Second, if the module is used in different places, all of them
places are impacted by the conditional use of the module.

It is better to create, for this module \code{M}, a new implementation
\code{M\_empty}, in which all the functions do nothing (or are
replaced by macros that do nothing)\footnote{Note that this much
  easier to achieve if the interface only expose functions, and not
  global variables; this is one of the reasons why it is preferable to
  hide global variables with static and use accessor/mutator
  functions.}. This leads to less configuration options, less code,
and code easier to read.


\subsection{Dividing configuration options}
\label{sec:divid-conf-opti}

Options should be split into the smallest possible pieces. One could
think that too much splitting of options would lead to a proliferation
of configuration options, and would make options dependencies
difficult to understand.

On the contrary, having many options and modules makes their meaning
more precise. Each configuration option names a concept of the
application, and giving names to precise concepts helps greatly in
their understanding. Moreover, it makes the system more modular.

As an example, the OASIS micro-kernel defines a \code{date} configuration option
that accepts three different values: \code{date16}, \code{date32}, and
\code{date64}, which sets the size of a \code{date\_t} integer
type. The original code was written with the assumption that 16 and 32 bits
\code{date\_t} may lead to a date overflow, whereas a 64 bit field may not;
therefore all the overflow-handling code was enclosed by \code{\#if defined(
DATE16) || defined(DATE32)} directives, which does not seem appropriate at first
glance.

We reworked this using \configen{}, and here is the result:

\medskip\hrule\smallskip
\begin{minipage}[htbp]{1.0\linewidth}
\begin{verbatim}
date -> date_size & date_overflows?
# (auto) date_overflows? : date_overflows_yes | date_overflows_no
date_size: date16 | date32 | date64
date16 -> date_overflows_yes; date32 -> date_overflows_yes
date64 -> date_overflows_no
\end{verbatim}
\end{minipage}
\smallskip\hrule\medskip

Even if we added new configuration options, the resulting code is
easier to understand, as the ``overflow'' concept is named and
assumptions are explicit. It is also more modular, as we could easily
change the code to allow 32 bit dates that do not overflow.

\xxxpaslaplace{Splitting modules also allows more code reuse. The OASIS
  kernel has many high-level functionalities, and works on many
  different hardware platforms. Splitting it into modules has helped
  reusing many low-level code that depends on low-level architectural
  features which was previously hard to reuse, and avoided restricting
  some high-level functionalities to only certain platforms.}

\xxxpaslaplace{Autre example: - Dire egalement que plus d'options de
  configurations de nuisent pas a la lisibilite du code, mais au
  contraire l'ameliore. La raison en est la suivante: si on limite le
  nombre d'options de configuration, on cree de "fausses dependances".
  I.e., on va dependre de "PowerPC | arm" plutot que de dependre de
  "LLSC". Il faut alors avoir que le powerpc et l'arm implementent
  llsc pour comprendre le code. Avec notre configuration, on depend
  juste de ce qu'il faut dependre, i.e. llsc; la visualisaion du
  graphe permet ensuite de savoir ce qui implemente llsc. cela
  ameliore a la fois la comprehension, la modularite, et minimise la
  taille de code requis en maximisant la reutilisation.}

\subsection{Testing code}

When writing a unit test for a module \code{M}, some code has to be
activated to test \code{M} (e.g. calls to M and checking of M results),
and some code has to be activated to satisfy M dependencies (for
instance, unit testing of our scheduler requires a special version of the
context switching functions, that only log context-switch operations).

So far we found \configen{} to be of great help to automate the
activation of these requirements. \xxxpaslaplace{, for all
  kind of testing (functional and integration tests, black box unit
  tests of interfaces, white box unit tests of implementations)}. 
However there is still some work to do to improve this automation.

\section{The \configen{} prototype}
\label{sec:configen-prototype}

\subsection{The \configen{} script}

Aside from the logic solver, \configen{} is an extremely simple tool,
composed of less than 400 lines of Ruby code. Yet this tool does parse
the \code{deps} file, implements the graph user interface, interacts with
the solver, and outputs the \code{config.h} and \code{config.mk}
files.

The parser, that builds the dependency graph from the \code{deps}
file, was really easy to write because our syntax defines a regular
language, and thus can be parsed easily using simple regular
expressions.

To avoid the tedious development of a complex HMI, we use a
\texttt{graphviz}\footnote{An open-source graph vizualization
  software: \url{http://www.graphviz.org}.} feature that can output images and
HTML image maps such that clicking on a node would send different HTTP
requests. So all \configen{} has to do is to output the graph to the
dot file format with the correct options, implement a small web server
to handle the different ``node clicked'' requests, communicate with
the logic solver, and ask graphviz to do all the redisplay work with
the result of the solver. This way, a standard web browser stands for the
graphical interface. Printing the \code{config.h} and \code{config.mk} files was
just trivial scripting.

\xxxpaslaplace{Still, this tool is quite young, and its code is not yet very clean.
We have often reworked the syntax or the graphical display, as we
found better ways to adapt it to our needs. \xxx{We are considering
  opening it once it becomes a bit more stable and mature.}}


\subsection{The logic solver}
\label{subs:logicsolver}

Every time the user clicks on a node, he sets the corresponding configuration
symbol (i.e. logical literal) to a truth value, sequentially \texttt{TRUE} (1),
then \texttt{FALSE} (0), then back to the ``unset'' state. The idea is, after
each click, to infer which configuration symbols \emph{have} to be \texttt{TRUE}
or \texttt{FALSE} subsequently to the user action.

\subsubsection{Formal definition of the problem}

Let us define the following notations:
\begin{itemize}
    \item $X=(x_1, \dots, x_n)$ is the set of literals defined in the
      \texttt{deps} file, and $f(x_1,\dots,x_n)$ the boolean expression
      corresponding to the dependency graph. 
    \item $\mathcal{A}$ is a boolean clause defining the \emph{partial truth
          assignment}, as defined by the clicks of the user. We note $U_1\subset
          X$ (resp. $U_0\subset X$) the set of literals forced to 1 (resp. to 0)
          by the user.
          For the rest of this section, we assume without loss of generality
          that the literals are ordered as follows:
          \[ \underbrace{x_1\ ,\ \dots\ ,\ x_m}_{= U_0}\ ,\ \underbrace{x_{m+1}\
          ,\ \dots\ ,\ x_p}_{=U_1}\ ,\  x_{p+1}\ ,\ \dots\ ,\ x_n
          \qquad\text{ with } m \leq p < n \]

          Therefore we have: $\mathcal{A} \equiv 
          \displaystyle\bigg (\bigwedge_{i\in \{1, \dots, m\}}\neg x_i \bigg )
          \wedge
          \bigg (\bigwedge_{j\in \{m+1, \dots, p\}} x_j \bigg )
          $

    \item Let $f_\mathcal{A} = f \wedge \mathcal{A}$. 
          I.e. $f_\mathcal{A}$ is the function obtained after setting in the
          expression of $f$ all the literals in $U_1$ and $U_0$.
\end{itemize}
Then our problem is to find the biggest subsets $S_0$ and $S_1$ of
$\{x_{p+1},\dots,x_n\}$ such that $S_0 \cap S_1 = \emptyset$ and:
\begin{eqnarray*}
    \forall x \in S_0,\qquad \big\{f_\mathcal{A} & \Rightarrow & \neg
        x\big\} \text{ is a tautology}\\
    \forall x \in S_1, \qquad \big \{f_\mathcal{A} & \Rightarrow & x\big\}
        \text{ is a tautology}
\end{eqnarray*}

\begin{theorem}
    \label{thm}
    The problem of finding whether a given literal is in $S_1$ (resp. $S_0$) is
    co-NP-complete.
\end{theorem}
Before proving this theorem, let us show the following result:
\begin{lemma}
  \label{lem:gen}
  The satisfiability problem $(P)$ of a boolean expression $f(x_1, \dots, x_n)$
  described by the \code{deps} file (whose operators are described in section
  \ref{subsubsec:basic-ops}) is NP-complete.
\end{lemma}
\begin{proof}[Proof of Lemma \ref{lem:gen}]
    Let us prove first that $(P)$ is a NP-hard problem. For that purpose, we can
    easily reduce any 3-SAT instance, a well-known NP-complete problem, to a
    formula such as $f$.  Indeed, each clause $(y_1 \vee y_2 \vee y_3)$ of a
    3-CNF expression can be written using the interface/implementations operator
    as: $\neg \big (0:(y_1 | y_2 | y_3)\big )$ (\emph{see the footnote on
    p.\pageref{foot:operators} for the expression of the ``$\neg$'' operator and
    the ``$0$'' Boolean constant with our operators}). As this transformation
    can clearly be processed in polynomial time, $(P)$ is then NP-hard. 
    
    But $(P)$ is also in NP. Indeed, an algorithm that
    non-deterministically chooses the Boolean value of each literal
    $(x_1, \dots, x_n)$ can easily decide in polynomial time if the
    formula $f(x_1, \dots, x_n)$ is true for the chosen valuation.

    Since $(P)$ is both NP and NP-hard, it is NP-complete.
\end{proof}

\begin{proof}[Proof of Theorem \ref{thm}]
    We will prove this theorem for literals in $S_1$; the case of literals
    belonging to $S_0$ is almost identical. 

    Provided a partial truth assignment $\mathcal{A}\equiv
          \left(\bigwedge_{i\in \{1, \dots, m\}}\neg x_i \right )
          \wedge
          \left (\bigwedge_{j\in \{m+1, \dots, p\}} x_j \right )$
    and an unvalued literal $x_k$ (with $k\in \{m+1,\dots,n\}$), let us note
    ($P'$) the problem of deciding if $\underbrace{f(x_1,\dots,x_n)\wedge
    \mathcal{A}}_{f_\mathcal{A}} \wedge \neg x_k$
    is satisfiable. Note that $(f_\mathcal{A}\Rightarrow x_k)$ is a tautology
    iff $f_\mathcal{A}\wedge \neg x_k$ is \emph{not} satisfiable. Therefore,
    proving that the complement problem $(P')$ is NP-complete will prove the
    theorem.    
    
    We can reduce in polynomial time any problem ($P$)
    to a ($P'$) problem by extending the set of literals addressed by $(P)$. Let
    $y\not\in X$ and $z\not\in X$ be any two literals, then the following
    formula is an instance of $(P')$ with $n+2$ variables:
    $$\underbrace{f(x_1,\dots,x_n) \wedge ( y \Rightarrow y ) \wedge ( z
      \Rightarrow z )}_{F(y, z, x_1,\dots,x_n)} \wedge
    \underbrace{y}_{\mathcal{A}} \wedge \neg z$$ With the previous
    notations, we have in this case $U_0 = \emptyset$, $U_1=\{y\}$,
    $\mathcal{A}=y$. If this formula is satisfiable, then so is $f$;
    conversely, if $f$ is satisfiable, the above formula is also
    satisfiable (with $y$ set to ``\emph{true}'' and $z$ set to
    ``\emph{false}''). $(P')$ is thus a NP-hard problem. For the same
    reasons than for $(P)$ (\emph{see Lemma~\ref{lem:gen}}), it is
    also NP-complete.

    Therefore, deciding if a literal belongs to $S_1$ is indeed a co-NP-complete
    problem.
\end{proof}

We have proved here that in the most general case, the problem addressed is
co-NP complete. However, deciding if $f_\mathcal{A}\Rightarrow x_k$ is a
tautology is meaningful only if $f_\mathcal{A}$ is satisfiable itself, i.e. if
the logical description of the system is ``coherent'', and if the options chosen
so far by the user are not contradictory.
Therefore, another approach would be to ensure this property first with a
regular SAT-solver, then to search if $f_\mathcal{A}\wedge \neg x_k$ is
satisfiable or not. This last step is probably easier than a co-NP problem.
Moreover, once $f$ has been proved to be satisfiable, it should be easier to
prove that $f_\mathcal{A}$ is satisfiable every time the user iteratively
appends new literals to $\mathcal{A}$ by clicking.

\subsubsection{Internals of the solver}

Our logic solver applies a simple and intuitive heuristic to compute subsets of
$S_0$ and $S_1$. The idea is to compute and simplify the $f_\mathcal{A}$
expression for each assignment $\mathcal{A}$ provided by the user, then to
convert it to a CNF form using a straightforward algorithm\footnote{The
algorithm consists in recursively applying distributivity property of the
$\wedge$ operator.}. Then, all clauses of the final expression that are literals
(resp. negated literals) belong to $S_1$ (resp. $S_0$).

To this purpose, the dependency graph expressed in \texttt{deps} is
translated into a boolean expression that only uses $\wedge$, $\vee$, and
literal $\neg$\ \footnote{By literal $\neg$ we mean that the operator may only be
applied to boolean variables, not to operators. E.g. $\neg (a \wedge b)$ is
prohibited, whereas $(\neg a)\vee(\neg b)$ is not.}. The formal simplifier can
then manipulate this boolean expression and apply the following basic logic
rules:
\begin{center}
\begin{tabular}{cccc}
    $x \wedge 1 = x$ & $x\wedge 0 = 0$ & 
      $x \wedge f(x, \dots) =x \wedge f(1,\dots)$ &
      $\neg x \wedge f(x, \dots) = \neg x \wedge f(0, \dots)$ \\
    $x \vee 1 = 1$ & $x\vee 0 = x$ & 
      $x \vee f(x, \dots) = x \vee f(0, \dots)$ &
      $\neg x \vee f(x, \dots) = \neg x \vee f(1, \dots)$ \\
\end{tabular}
\end{center}
before converting the result to a CNF. The literals and negated literals
clauses are then extracted and sent back to the Ruby script for display.

In most cases, our solver managed to find the whole subsets $S_0$ and
$S_1$. In some however, it failed to see all dependencies. For instance, in the
example of figure \ref{fig:example-graphical-representation}
p.\pageref{fig:example-graphical-representation}, our solver was actually unable
to deduce from $\{\text{\texttt{arm}}=1,\text{\texttt{sched}}=1\}$ that
\texttt{llsc} (and subsequently \texttt{llsc\_arm}) is always
true.\footnote{Indeed, the mandatory choice of one option in \texttt{clock} as
well as in \texttt{ctxlist} will either set \texttt{llsc} directly, or will set
\texttt{spinlock}, then \texttt{spinlock\_llsc} which is the last available
choice, then \texttt{llsc}.} 
The reason for this failure is that the policy described above is not sufficient
to deduce from:
$$( a \oplus b ) \wedge ( a \Rightarrow c ) \wedge ( b \Rightarrow c )
  \quad\equiv\quad
  ( a \vee b)\wedge(\neg b \vee \neg a)\wedge(\neg a \vee c)\wedge(\neg b \vee
  c)$$
that $c$ is necessarily always true.

Even if the solver shows its limitations, it remains correct in the sense that
it will never deduce an erroneous literal value, e.g. that would define unwanted
options, or that would result in an unsatisfiable expression. 

The solver is written in C; its performances were not measured precisely, but
for all the dependencies trees that we used so far to model the OASIS kernel,
the calculus time appeared immediate. It has not been tested yet on larger scale
projects, mainly because it is still an ``ad-hoc'' tool, that requires a lot of
improvements before being subject to a relevant performance evaluation.

Evolutions will be discussed in section \ref{sec:conclusion}. Although the
solver approach is obviously not suitable for (even approximate) solving of
SAT-problems, (especially when compared to dedicated tools such as MiniSat
\cite{bib:minisat}), we believe we can make it more efficient by focusing only
on a meaningful restricted set of boolean expressions, e.g. only those
represented by an acyclic graph.


\section{Related works}
\label{sec:state-art}

Using boolean logic to manage and validate complex dependencies schemes has been
done before in different application domains, including software architecture.

Our development approach is close to the \emph{Software Product Line}
engineering technique \cite{bib:lee}, as it promotes modularity and
re-usability as development-driving key concepts. The features of a
SPL are usually represented as an oriented graph (\emph{feature
  diagram}). The semantics of this graph has been formalized and
studied \cite{bib:bontemps}, although to the best of our knowledge it
is not used in any practical application.

The \emph{Kconfig} Linux kernel configurator is a tool similar to ours.  With
the use of \emph{Kconfig} script files and a dedicated syntax, the kernel
developers have a powerful and efficient way to express internal dependencies.
The user has therefore a great freedom in the choice of his kernel components
(see Sincero's work \cite{bib:SinceroS08} -- an attempt to bridge the gap
between the SPL and the Open Source communities, and the Linux Kernel
documentation\footnote{\url{http://www.kernel.org}, see
\texttt{Documentation/kbuild/} directory}). However no graph representation of
the dependencies is provided, nor any interactive interface such as ours.

The main link between software components dependencies and propositional logic
comes from the work of Mancinelli, Abate, Boender and Di Cosmo on Free
Open-Source Software distributions, through the EDOS and Mancoosi projects
\cite{bib:edos}. In \cite{bib:mancinelli}, a SAT-solver is used to address the
installability problem of a set of packages; in \cite{bib:abate}, the dependency
graph of a package repository is analyzed to identify ``sensible'' components
that may widely impact the system if corrupted or removed. 

\section{Conclusion}
\label{sec:conclusion}

This paper has presented \configen{}, a tool for managing software configuration
options. We exposed the concepts behind \configen{}, showed how it can be
integrated in a software development project, and described a set of good
practices and examples that come from our experience using \configen{} for
system development. We also presented the graphical interface of \configen{},
the associated logic solver and the theoretical problem it addresses.

\configen{} is still a prototype, but has already proved to be very
useful. The tools are simple to use and have helped in refactoring
a complex software, making it easier to understand. It also encourages
good software practices. We found the graphical interface of great help when
defining new modules.

There are many future possible developments. It might be interesting to extract
the dependency file from the source code, using source code annotations for
instance. Another interesting point would be to guide the user's choices with
``automatic'' implementations, that would discard by default rarely used options
(e.g. benchmarking modules).
At last, the work of \cite{bib:abate} could also be used to isolate critical
features, with application to quality assurance.

Many interesting developments also remain to be done in the solver.
The problem we need to solve is co-NP-complete in the general case;
however we did not take into account many restrictions yet. For
instance, the current proof uses implementation/interfaces operators
in which some implementations belong to multiple interfaces, which
does not happen in real use. Moreover, we do not use cycles in the use
cases encountered so far. It is possible that with such restrictions,
the problem we try to solve becomes polynomial. Even if it is not,
there are strong relationships between successive iterations of the
problem to solve (i.e. they differ by only one truth assignment),
which could be exploited by an incremental solver. Meanwhile, it would
be more appropriate to connect to a SAT-solver and get the complete
solution, as suggested in \ref{subs:logicsolver}. Such solvers could
also be used to get the set of all the possible configurations, e.g.
for testing purposes.


\renewcommand{\baselinestretch}{0.90}

\vspace{-3mm}
\nocite{*}
\bibliographystyle{eptcs}
\bibliography{lococo}

\end{document}